% Preamble
\def\ovarphi{{\varphi^{\kern-3.5pt \raise 2 pt
\hbox{$\scriptscriptstyle 
0$}}}{} }
\def\og{{g^{\kern-4pt \raise 1.5 pt\hbox{$\scriptscriptstyle
0$}}} {} }
\font\bigbf=cmbx10 scaled\magstep2
%
%    Line spacing
%

\def\oneandathirdspace{\baselineskip=\normalbaselineskip
  \multiply\baselineskip by 4 \divide\baselineskip by 3}
\def\oneandahalfspace{\baselineskip=\normalbaselineskip
  \multiply\baselineskip by 3 \divide\baselineskip by 2}

\overfullrule=0pt
\magnification=1200
\oneandahalfspace
\nopagenumbers
\line{\hfill May 1997 }
\vskip 1truein
\centerline{\bigbf LOCAL COHOMOLOGY IN FIELD THEORY}
\centerline{\sl with applications to the Einstein equations}
\vskip 0.6truein
\centerline{C. G. Torre}
\smallskip
\centerline{\it Department of Physics}
\centerline{\it Utah State University} 
\centerline{\it Logan, UT 84322-4415, 
USA}
\centerline{\tt torre@cc.usu.edu}
\vskip 2truein

\noindent
This is an introductory survey of the theory of $p$-form conservation
laws in field
theory.  It is based upon a series of lectures given at the Second
Mexican School on Gravitation and Mathematical Physics held in
Tlaxcala, Mexico {}from December 1--7, 1996.  Proceedings available online
at http://kaluza.physik.uni-konstanz.de/2MS/ProcMain.html.
\vfill\eject
\count0=1
\footline={\hss\tenrm\folio\hss}

\noindent{\bf 1. Introduction.\hfill}

In these lectures I will provide a survey of some aspects
of the theory of local cohomology in field theory and
provide some illustrative applications, primarily taken
{}from the Einstein equations of the general theory of
relativity.  There are many facets to this branch of
mathematical physics, and I will only concentrate on a
relatively small portion of the subject. A sample of some
of the recent literature on the subject can be found in [1].

For the purposes of these lectures I will restrict the
development of local cohomology to the theory of $p$-form
conservation laws for field equations.  This theory can be
considered to be a generalization of Noether's theory of
conserved currents to differential forms of any degree. 
Rather than presenting the theory in its full generality, I
will illustrate each point via examples taken {}from
relativistic field theory, with the principal example being
the vacuum Einstein equations.  In so doing I will summarize the
results of a
classification of all $p$-form conservation laws that
can be locally constructed {}from an Einstein metric and its
derivatives to any finite order [2].  Finally, I show how the theory
and techniques used to analyze $p$-form conservation laws
can be used to give a useful derivation of asymptotic
conservation laws ({\it e.g.}, ADM energy), when they exist, for
any field theory.  

Many of the basic results on $p$-form conservation laws can be
obtained {}from a variety of points of view [1].  Virtually all of the
results
presented here were
obtained in collaboration with Ian Anderson.  We are
currently preparing an 
in-depth exposition of a theory of $p$-form conservation laws [2],
relative
to which these lectures can be considered an introductory survey.

An outline of the topics to be covered is as follows:
\medskip
\item{$\bullet$} $p$-form
conservation laws in field theory:  introduction and examples.

\item{$\bullet$}  Jet bundle of metrics and the
Euler-Lagrange complex.

\item{$\bullet$}  Jet bundle of Einstein metrics.

\item{$\bullet$}  $p$-form conservation laws and local
cohomology.

\item{$\bullet$}  Identically closed $p$-forms: cohomology of the free
Euler-Lagrange complex.

\item{$\bullet$}  Identically closed $p$-forms locally
constructed {}from a metric:  gravitational kink number.

\item{$\bullet$}  $(n-1)$-form conservation laws and
infinitesimal symmetries of the field equations.

\item{$\bullet$}  Lower-degree conservation laws and
infinitesimal gauge symmetries of solutions to the field
equations.

\item{$\bullet$}  Asymptotic conservation laws and local cohomology.
\medskip

In order to give a reasonably clear picture of the theory without
getting mired in technical details, my presentation will be rather
informal and lacking in rigor. 
The reader
who wishes to see a more careful and rigorous development of this
material can
consult the references in [1] and the upcoming publication [2], which
I hope should be
completed
around the time these lecture notes become available.
\bigskip

\noindent{\bf 2. $p$-form conservation laws in field theory: 
introduction and examples.\hfill}

Let $M$ be a differentiable manifold of dimension $n$ and suppose we
are given a
differential $p$-form, $p<n$, on $M$. All geometric objects that we
shall use are
considered to
be (as) smooth (as necessary).  To any $p$-dimensional closed
submanifold
$\Sigma\hookrightarrow M$, one can associate a real number
$Q(\Sigma)$ by
integration:
$$
Q(\Sigma)=\int_\Sigma \omega.
$$
In general, the number $Q(\Sigma)$ is not particularly interesting
since
it depends on the arbitrary choice of $\Sigma$.  However, if $\omega$
is a
{\it closed} $p$-form,
$$
d\omega = 0,
$$
then it follows that $Q(\Sigma)$ is unchanged by continuous
deformations
of 
$\Sigma$.  Conversely, $Q(\Sigma)$ is unchanged by  continuous
deformations
of 
$\Sigma$ only if $\omega$ is a closed form.  More precisely,
$Q(\Sigma)$
depends
only on the homology class of $\Sigma$ if and only if $\omega$ is a
closed
form.  Of course, if $\omega$ is an {\it exact} form, that is, there
exists
a $(p-1)$-form $\alpha$ such that
$$
\omega=d\alpha,
$$
then because $d^2=0$ it follows that $\omega$ is closed, but by
Stokes' theorem
it follows that $Q(\Sigma)=0$ for any choice of
$\Sigma\hookrightarrow M$.  Thus one can use the equivalence classes
of
closed
modulo exact forms to study the homological properties of $M$, and
this is
the basis of de Rham cohomology.  All of these notions can be
generalized
to open manifolds (that is, those with asymptotic regions) and/or
manifolds with boundary provided one imposes appropriate boundary
conditions.

In field theory one, in effect, uses de Rham cohomology classes to
represent 
{\it
$p$-form conservation laws}.  Roughly speaking, a $p$-form
conservation
law is an assignment of a closed spacetime $p$-form $\omega[\varphi]$
to each solution
$\varphi$
of
the field equations. Given a $p$-form conservation law, one can
associate a number
$Q[\varphi]$ to each solution of the field equations by integrating
the
$p$-form over an appropriate $p$-dimensional submanifold $\Sigma$. 
Because
the resulting number is unchanged by continuous deformations of the
embedding of $\Sigma$ in spacetime, we say that the number is a
``conserved quantity''.  Indeed, when $\Sigma$ is, or is contained
in, a
spacelike hypersurface, the invariance of $Q$
with
respect to continuous deformations of $\Sigma$ corresponds to the
familiar
notion of a ``constant of the motion''.  One way to make a $p$-form
conservation law for any field theory is to associate an exact
$p$-form to all
solutions of the field equations.  Of course, in this case
$Q[\varphi]=0$ for
all solutions.  Consequently, we say that such $p$-form conservation
laws are
{\it trivial}.  We are therefore interested in cohomology classes,
that is,
equivalence classes of maps {}from the space of field configurations
into closed
modulo exact $p$-forms on spacetime.  In this sense the study of
$p$-form
conservation laws is a branch of local cohomology in field theory. 
Despite
their similarities, local cohomology in field theory is rather
different than
de Rham cohomology of manifolds.  The de Rham cohomology involves
equivalence
classes of
closed forms modulo exact forms on a manifold, while the local
cohomology studies equivalence
classes of
maps {}from field configurations into closed spacetime $p$-forms.  In
effect,
local cohomology in field theory involves a theory of {\it local
formulas}
for defining closed forms {}from spacetime field configurations.  The
de Rham cohomology detects global properties of
manifolds; in the small, the de Rham cohomology is always trivial. 
Local cohomology can be non-trivial even when one works in the small.
%
%Extra material
%

At this point it is worth making clear the way we are using
the word ``local''.  By far the most common use of the term is the
one which signifies ``in the small'' or ``in a sufficiently small
neighborhood''.  This is {\it not} the use of the word in ``local
cohomology'', or in ``$\ldots$ locally constructed {}from the
field...''.
This use of the word ``local'' signifies that the objects under
consideration depend, at a given point $q$ in spacetime, upon the
fields and their derivatives to some finite order at
$q$.  Thus, the Einstein Lagrangian density $\sqrt{|g|}R$ is a ``local
density'' in the sense that it is a scalar density which is
constructed locally {}from the metric and its first two derivatives. 
Local here does not mean, of course, in the small since the
Lagrangian density is globally well-defined.  By contrast, the value
of a vector at a point $p$ obtained by parallel propagation of a
given vector at a point $q$ is {\it not} a local metric quantity in
the sense in which we want to use the word ``local''.  This is, of
course, because the value of the vector at $p$ depends upon the
metric and its first derivatives along the curve connecting $q$ and
$p$.  Thus the connotation of local in ``local cohomology'' is not
that the considerations are only valid in the small, but rather that
the objects being used depend in a local fashion on the fields of
interest.

The above notion of $p$-form conservation laws will become
considerably
clearer if we introduce some examples.
\bigskip
\noindent{\sl 3-form conservation law:  charged scalar field in 4
dimensions.\hfill}

Consider a complex scalar field $\varphi$ on a 4-dimensional
spacetime $(M,g_{ab})$
satisfying the Klein-Gordon equation
$$
\nabla_a\nabla^a\varphi-m^2\varphi=0.\eqno(2.1)
$$
We define the 3-form
$$
\omega = \epsilon_{abcd} J^d dx^a \wedge dx^b \wedge dx^c\eqno(2.2)
$$
where
$$
J^d=i(\varphi^*\nabla^d\varphi-\varphi\nabla^d\varphi^*).
$$
The form $\omega$ and the vector field $J^d$ are locally constructed
{}from the Klein-Gordon field in the sense that at a given point they
depend upon only 
$\varphi$ and $\nabla_a\varphi$ {\it at that point}.  It is a nice
exercise to check that the exterior derivative of
$\omega$
vanishes by virtue of the field equation (2.1) and its complex
conjugate.  This is equivalent to the statement that the current
$J^a$ is
divergence-free ``on-shell'', which is very straightforward to verify
{}from
the identity:
$$
\eqalign{
\nabla_d J^d &=
i(\varphi^*\nabla^d\nabla_d\varphi-\varphi\nabla^d\nabla_d\varphi^*)
\cr
&=i[\varphi^*(\nabla_a\nabla^a\varphi-m^2\varphi)-
\varphi(\nabla_a\nabla^a\varphi^*-m^2\varphi^*)].}
$$
With appropriate boundary conditions, the integral of $\omega$ over a
spacelike hypersurface defines a conserved $U(1)$ charge for the field
theory which can be coupled, {\it e.g.}, to an electromagnetic
field.  

In
field theory conservation laws usually arise via conserved 3-forms in
4 spacetime dimensions (or conserved $(n-1)$-forms in $n$
dimensions), such as in the example above.
But as
the following examples show, there are also a variety of
important ``lower-degree''
conservation laws.

\bigskip

\noindent{\sl 2-form conservation law: electromagnetic field.\hfill}

The electromagnetic field can be described by an antisymmetric tensor
field, the ``field strength tensor'' $F_{ab}$, built {}from a one-form
$A_a$
via
$$
F_{ab} = \nabla_a A_b - \nabla_b A_a.
$$
In regions of spacetime not containing electromagnetic sources the
field
strength satisfies the field equations
$$
\nabla^b F_{ab}=0.\eqno(2.3)
$$
In 4 spacetime dimensions, a 2-form conservation law for these field
equations is provided by
$$
\omega = \epsilon_{abcd} F^{cd} dx^a\wedge dx^b.\eqno(2.4)
$$
It is a straightforward exercise to verify that $\omega$ in (2.4) is
closed
when the field equations (2.3) are satisfied.  The integral of
$\omega$
over
a spacelike  two-sphere yields the electric charge contained in the
two-sphere. This is, of course, just a version of Gauss' law, which we
see can
be interpreted as a 2-form conservation law.

\bigskip

\noindent{\sl 2-form and 1-form conservation laws: vacuum spacetimes
with
a
Killing vector.\hfill}

Consider vacuum regions of spacetime satisfying the Einstein equations
$$
G_{ab}=0,\eqno(2.5)
$$
and which admit a Killing vector field, that is, a vector field $K^a$
such
that
$$
\nabla_{(a} K_{b)} = 0.\eqno(2.6)
$$ 
Associated with these equations there is a 2-form conservation law,
the
Komar 
2-form, 
$$
\kappa=\epsilon_{abcd}\nabla^cK^d dx^a\wedge dx^b,\eqno(2.7)
$$
and a 1-form conservation law, the ``twist'' 1-form,
$$
\tau= \epsilon_{abcd}K^b\nabla^cK^d dx^a.\eqno(2.8)
$$
To verify that these are indeed conservation laws, simply take their
exterior derivative and you will find, with a little algebra, that the
result is algebraic in the Einstein tensor $G_{ab}$, the Lie
derivative
${\cal L}_Kg_{ab}$, and its derivative $\nabla_c{\cal L}_Kg_{ab}$.  
The integral of the Komar 2-form over a closed
spacelike surface, {\it e.g.}, a sphere in the vacuum region
surrounding a
star, defines conserved energy, momentum, or angular momentum,
depending
upon the nature of the Killing vector field.  
\bigskip

\noindent{\sl 0-form conservation law: dilaton gravity in two
dimensions.\hfill}

Here the spacetime is two-dimensional and the fields are a metric
$g_{ab}$
and a scalar field $\varphi$.  Using the derivative $\nabla$
compatible
with
the metric, the field equations are
$$
R[g]+\lambda^\prime(\varphi)=0,\eqno(2.9)
$$
and
$$
\nabla_a\nabla_b \varphi - {1\over2}
\lambda(\varphi) g_{ab}=0,\eqno(2.10)
$$
where $R[g]$ is the scalar curvature of $g_{ab}$ and
$\lambda(\varphi)$
is a
local function of $\varphi$. 
These
field equations admit a 0-form conservation law
$$
\alpha=\nabla^a\varphi\nabla_a\varphi -\int d\varphi \lambda(\varphi).
\eqno(2.11)
$$
The 0-form $\alpha$ is a local
function of the fields and their first derivatives which becomes a
constant when the field
equations are
satisfied.
This is easily verified by computing
$$
\nabla_a\alpha=2(\nabla_a\nabla_b \varphi - {1\over2}
\lambda(\varphi) g_{ab})\nabla^b\varphi.
$$
The on-shell (constant) value of $\alpha$ corresponds, {\it e.g.}, to
the
``mass'' of the dilatonic black hole [3].  The existence of a 0-form
conservation law is somewhat remarkable in a field theory since such
conservation laws normally arise as integrals of motion in mechanical
systems with a finite number of degrees of freedom (which can be
thought
of as field theories in one spacetime dimension).  The conservation of
$\alpha$ in dilaton theories reflects the fact that the pure dilaton
gravity theory (with no matter couplings) has only a finite number of
degrees of freedom once the action of the gauge group (spacetime
diffeomorphisms acting by pull-back on the the fields) has been
factored
out.  Indeed, up to a diffeomorphism, the fields $\varphi$ and
$g_{ab}$
solving (2.9), (2.10) can be obtained by solving a system of ordinary
differential equations for which the 0-form conservation law plays
the role of Hamiltonian.  In other words, the dilaton field equations
are,
modulo gauge, equivalent to a finite-dimensional dynamical system.
This result can be understood {}from a simple counting of constraints
in the Hamiltonian formulation of the theory.  There are 2 
first-class
 constraints associated with the field equations (2.9) and
(2.10).  Because there are four fields in the theory,
$(g_{ab},\varphi)$, there are no field degrees of freedom left over
after factoring out the gauge transformations.
\bigskip

\noindent{\sl Asymptotic conservation laws: the ADM energy.\hfill}

Our final example is the ADM energy, which is an instance of an {\it
asymptotic
conservation law},
that is, a $p$-form whose integral in an asymptotic region is
conserved by
virtue of field equations and (asymptotic) boundary conditions. 
The ADM energy in general relativity is defined
via a surface integral at infinity in an asymptotically flat spacelike
hypersurface [4].  One expression of it is
$$
E_{ADM} = {1\over16\pi}\lim_{r\to\infty}\sum_{i,j=1}^3
\int_{t,r={\rm
const.}}\left(g_{ij,j}-g_{ii,j}\right)d^2S^j.\eqno(2.12)
$$
Here the indices $i$ and $j$ refer to an asymptotically Cartesian
coordinate
chart $(t,x^i)$ on an asymptotically flat spacelike hypersurface
$t={\rm
constant}$.  The
coordinate $r$ is
the associated (asymptotic) radial variable and the integral is over a
coordinate 2-sphere with coordinate area element $d^2S^j$.  Given
asymptotically flat boundary conditions on the spacetime metric, it
can be
shown that $E$ is independent of the choice of $t$ and of the choice
of asymptotically Cartesian coordinates
whenever
the metric solves the vacuum Einstein equations.

This final example is not, strictly speaking, a $p$-form conservation
law
of the same type as those presented above, {\it e.g.}, because it is
not
displayed as a closed spacetime differential form.  However, we shall
show later
that it can nevertheless be understood using elements of our
theory of
$p$-form conservation laws.
\bigskip

Hopefully, these examples demonstrate that $p$-form conservation laws
are
ubiquitous and important in field theory.  The basic purpose of this
course is to give an introduction to a subset of techniques that have
been
developed recently to (i) explain the existence of these conservation
laws
field theoretically, and (ii) systematically find these conservation
laws,
when they exist, for any field theory.  Of course, for the
conservation
laws associated with conserved currents (e.g., 3-form conservation
laws in
4-dimensional spacetime), point (i) is addressed via the Noether
theory
relating symmetries of a Lagrangian to conservation laws, but even for
conserved currents results on point (ii) are probably less familiar
to
you (though they are well-developed, see [19]).  For lower-degree
conservation laws, points (i) and (ii) are only now being developed by
mathematicians and physicists [1,2].  In essence, my goal here is to
describe
machinery that allows one to find a generalization of Noether theory
to all 
$p$-form
conservation laws.

I would like to make this course a relatively accessible
introduction
to the theory of $p$-form conservation laws. Since the predominant
background of the participants of this school is in general
relativity, I
will
eschew a general, abstract treatment and develop the ideas largely in
the
context of a familiar and important example:  the vacuum field
equations
of Einstein's general theory of relativity.  

Our first step is to
give a
more precise characterization of what we mean when we say
``$p$-form
conservation law'', which formalizes the basic properties of
the
examples we have just discussed.  This is most easily done using the
language of jets.
\bigskip

\noindent{\bf 3. Jet bundle of metrics and the
Euler-Lagrange complex.\hfill}

The mathematical setting for the theory of $p$-form conservation laws
in a
field theory for a metric is the {\it jet bundle of metrics} $\cal
J$. 
Given a
spacetime manifold $M$ and local coordinates $x^i$, a point $p\in
{\cal J}$
is
specified by giving a spacetime point $x_0^i$, a metric at that point
$g_{ij}$ (a symmetric, non-degenerate rank-two tensor with the
appropriate
signature), and a sequence of tensors $g_{ij,k}$, $g_{ij,kl}$,
$\ldots$
representing the value of the derivatives of the metric at $x_0^i$. 
We
write
$$
p=(x_0, g_{ij}, g_{ij,k_1}, \dots, g_{ij,k_1\dots
k_n},\dots)\in{\cal J}.\eqno(3.1)
$$
In this way a local coordinate chart on $M$ defines a chart on ${\cal
J}$. 
It can
be
shown that an atlas on $M$ then leads to a well-defined global
construct,
namely ${\cal J}$, which is suitably independent of the choice of
atlas. 
Then,
as you might expect, it is possible to give an intrinsic,
coordinate-free
definition of ${\cal J}$.  See [5,19] for details.  For
simplicity, we
will
be content with a purely local, coordinate-based treatment, which
will be
adequate for the level of presentation of these lectures.  The space
${\cal J}$
so-constructed is a fiber bundle ${\cal J}\rightarrow M$. ${\cal J}$
can be viewed as a bundle ${\cal J}\rightarrow E$, where
$E\rightarrow M$ is the {\it bundle of metrics}.  Given a chart
on
$M$, a point in $E$ is specified by giving the pair $(x_0^i,
g_{ij})$.  A
metric tensor field $g_{ij}(x)$ defines a cross-section of both $E$
and ${\cal J}$. 
Conversely,
given a
point $p$ in ${\cal J}$ (or $E$), one can always find a smooth metric
tensor
field which takes the values defined by $p$.

It is often convenient to label points in ${\cal J}$ in a way that is
better
adapted to the geometric meaning associated to the metric and its
derivatives [6].
To this end, let $\Gamma^i_{jk}$ be the Christoffel symbols of the
metric 
$g_{ij}$, and let us recursively define ``prolonged Christoffel
symbols'', which are to be viewed as functions on the jet bundle, 
by
$$
\Gamma^i_{j_0j_1\cdots j_k}=\Gamma^i_{(j_0j_1\cdots j_{k-1},j_k)}
-(k-1)\Gamma^i_{m(j_1\cdots j_{k-2}}\Gamma^m_{j_{k-1}j_k)}, \quad 
k=1,2,\ldots\ \ .\eqno(3.2)
$$
I will schematically denote these variables by $\Gamma^{(k)}$
($\Gamma^{(1)}$ denotes the usual Christoffel symbol).  The 
$\Gamma^{(i)}\ i=1,\dots,l$ are algebraically independent at any
given spacetime 
point, and capture all of the spacetime coordinate information that
is hiding 
in the first $l$ derivatives of the metric.  Roughly speaking, given
metric components $g_{ij}(x)$ in some chart, the
$\Gamma^{(k)}$ 
measure the difference between the given coordinates $x^i$ on
spacetime 
and a geodesic coordinate system for the metric $g_{ij}(x)$.  In
particular, all 
of the $\Gamma^{(k)}$ variables vanish at the origin of a geodesic 
coordinate system.  Next, let us examine the spacetime geometric
content of 
the jet bundle.  Of course we expect geometrical information to enter 
through the curvature tensor and its covariant derivatives. But these 
quantities do not give a good parametrization of the jet bundle
because they 
are subject to the Ricci and Bianchi identities as well as identities
obtained 
{}from these by differentiation, and so these variables are not
algebraically independent at a given point in spacetime.  To handle
these identities we
introduce the 
following tensors, which are being viewed as functions on the jet
bundle,
$$
Q_{ij},_{j_1\cdots 
j_k}=g_{ir}g_{js}\nabla_{(j_3}
\cdots\nabla_{j_k}
R^{r\phantom{(j}s}_{\phantom{r}j_1\phantom{s} j_2)}, \quad
k=2,3,\ldots\ \ .\eqno(3.3)
$$
Here $R_{abcd}$ is the Riemann tensor.  These tensors have the
symmetries
$$
Q_{ij},_{j_1\cdots j_k}= Q_{(ij)},_{j_1\cdots j_k}
= Q_{ij},_{(j_1\cdots j_k)},
$$
and
$$
Q_{i(j},_{j_1\cdots j_k)}=0,
$$
and can be shown to be algebraically independent at any given point
of spacetime.  
We will schematically denote these variables by $Q^{(k)}$. Note that
the variables $Q^{(i)},\ i=2,\dots,l$ depend upon the first $l$
derivatives of metric.  The variables $Q^{(k)}$
contain all 
spacetime-geometric information defined by a point in the jet
bundle.  
These variables were used by Penrose [7] and are closely related to 
Thomas's ``normal metric tensors'' [8].  It can be shown that the
variables
$$
(x^i, g_{ij},
\Gamma^{(1)},\Gamma^{(2)},\ldots,Q^{(2)},Q^{(3)},\ldots)\eqno(3.4)
$$
uniquely parameterize points in the jet bundle [6].  Put differently,
the
variables (3.4) are the freely specifiable data at a point of a
(pseudo-)Riemannian manifold.

The reason for introducing the notion of a jet bundle is that it is a
natural arena for building local functions of the metric and its
derivatives. So, for example, the scalar curvature of a metric,
$$
R[g]=R(g_{ij}, g_{ij,k}, g_{ij,kl}),
$$
can be
viewed in an obvious way as a map {}from ${\cal J}$ into the real
numbers. 
More
importantly for our purposes, we can define a {\it $p$-form locally
constructed {}from the metric}, $\omega[g]$ as a map {}from ${\cal J}$
into the (bundle of)
$p$-forms on $M$ depending upon an arbitrary but finite number of
derivatives of $g_{ij}$.   We write
$$
\omega[g]=\omega_{i_1\dots i_p}(x,g,\partial g,\cdots,
\partial^kg)dx^{i_1}
\wedge\cdots \wedge dx^{i_p}.
$$
As you can easily see, all of our previous examples of $p$-form
conservation
laws, in particular, the Komar and twist conservation laws, can be
viewed
as such maps {}from an appropriate jet bundle into the spacetime
$p$-forms. More generally, a tensor field $T[g]$ locally constructed
{}from the
metric is a map {}from ${\cal J}$ into the relevant bundle of tensor
fields.  Given a specific metric tensor field, that is, a cross
section of the bundle of metrics $E$, we have a corresponding cross
section of $\cal J$ which can be used to pull back $T[g]$ to a tensor
field on $M$. This fancy sounding construction simply means that we
substitute a given metric into the formula for $T[g]$.

Next we need to introduce an appropriate notion of derivatives of
functions on jet space.  To begin, let
$f[g]$ be a
0-form locally constructed {}from the metric, that is, $f[g]$ associates
a number to
each point in ${\cal J}$.  Again, the scalar curvature of the metric
is an
example of
such an object.  We define the {\it total derivative}, $D_if$, of
$f[g]$ using
the chain rule:
$$
D_if = {\partial f\over\partial x^i} 
+ {\partial f\over \partial g_{kl}}g_{kl,i} + \dots
+ {\partial f\over \partial g_{kl,i_1\dots i_p}}g_{kl,i_1\dots i_p i}
+\dots\ \ .
$$
This is the field-theoretic analog of the ``total time derivative''
in particle
mechanics.  Given a metric tensor field, $g_{ij}(x)$, we have a
cross section
$\sigma\colon M\to{\cal J}$;
the total
derivative is defined so that
$$
(D_i f)\circ\sigma={\partial\over\partial x_i}(f\circ\sigma),
$$
or, in an alternative notation,
$$
\Big((D_i f)[g]\Big)_{g=g(x)}={\partial\over\partial x_i}(f[g(x)]).
$$
Note that the total derivative of a function locally constructed {}from
the first $k$ derivatives of the metric is a function locally
constructed {}from the first $k+1$ derivatives of the metric.  We can
now define the (total) exterior derivative of a $p$-form
$\omega[g]$
locally constructed {}from the metric:
$$
D\omega[g] = D_{[k}\omega_{i_1\dots i_p]}(x,g,\partial g,\dots,
\partial^lg)
dx^k\wedge dx^{i_1}
\wedge\dots dx^{i_p}.
$$
Evidently, $D\omega$ is a $(p+1)$-form locally constructed {}from the
metric.  It
is a simple exercise to check that $D^2=0$.  Thus the spaces
$\Omega^p({\cal J})$ of
$p$-forms locally constructed {}from the metric form a {\it complex},
analogous
to the de Rham complex.  If the spacetime dimension is $n$, then
there will be
no non-zero $p$-forms with $p>n$.  In particular, if $\omega[g]$ is an
$n$-form, then $D\omega\equiv0$.  It would appear that the complex
thus
stops at
$\Omega^n({\cal J})$.  But it turns out that there is a useful
(infinite)
continuation of
the spaces 
$\Omega^p({\cal J})$ which defines the {\it Euler-Lagrange complex}
[9].
This complex gets its name {}from the fact that the next operator in
the sequence defines the Euler-Lagrange equations {}from a Lagrangian
(viewed as a 4-form).
%
%  Extra material
%

Although we shall not really need to use the full Euler-Lagrange
complex, I can briefly expand a little on the basic ideas.  To begin,
an $n$-form $\lambda[g]$ locally constructed {}from the metric
can be
viewed as defining a Lagrangian; the corresponding action functional
$S$ is
obtained via an integral over an $n$-dimensional region ${\cal
R}\subset M$ of spacetime:
$$
S=\int_{{\cal R}}\lambda[g].
$$
Let us extend our space of differential forms to include differentials
of the
field variables $(g_{ij}, g_{ij,k},\dots)$ ({\it i.e.}, we consider
differential forms on $\cal J$). The
Euler-Lagrange
equations $E^{ij}(\lambda)$ are obtained by applying a differential
operator
on ${\cal J}$---the
Euler-Lagrange
operator $E^{ij}(\cdot)$---to the Lagrangian.  The resulting
collection of functions of the metric and its derivatives defines an
$(n+1)$ form on
${\cal J}$:  
$$
E(\lambda)=E^{ij}(\lambda)dg_{ij}.
$$
We
thus can
define a new space $\Omega^{n+1}({\cal J})$.  The image of the
Euler-Lagrange
operator on
$n$-forms defines elements of $\Omega^{n+1}({\cal J})$, but this
operator is
neither
surjective or injective.  In particular, if $\alpha[g]$ is an
$(n-1)$-form, then
$D\alpha$ is
an $n$-form, {\it i.e.}, a Lagrangian, whose Euler-Lagrange
expression is trivial: $E^{ij}(D\alpha)\equiv0$.  This is just the
differential form expression of
the
familiar result that the Euler-Lagrange equations are trivial if the
Lagrangian
density is a total divergence.  Thus a non-trivial extension of the
operators
$$
D\colon\Omega^p({\cal J})\to \Omega^{p+1}({\cal J}),
$$
satisfying $D^2=0$, to the case where $p=n$ is to replace $D$ with the
Euler-Lagrange operator acting upon $\Omega^n$.  It is this element
of the
sequence that gives the Euler-Lagrange complex its name.  Since we
will only be
discussing the $p$-form conservation laws, the extension of the
complex for
$p$-forms of degree $p\geq n$ will not interest us.  However, this
extension is
quite important in field theory.  For example, the operator following
the Euler-Lagrange operator in the sequence annihilates all
$(n+1)$-forms that are in the image of the Euler-Lagrange operator. 
The cohomology at this form degree can then be profitably used to
study
the
inverse problem in the calculus of variations [9].
%
% end extra
%

We now can give a more precise definition of a ``closed $p$-form
locally
constructed {}from the metric''.  It is simply an element 
$\omega[g]\in\Omega^p({\cal J})$ satisfying $D\omega=0$.  However, we
have
not yet
injected into the discussion the idea that the $p$-form should be
closed when
the metric satisfies the field equations.  To do this we need to find
out what
it means to go ``on-shell'' using the jet bundle.

\bigskip

\noindent{\bf 4. Jet bundle of Einstein metrics.\hfill}

Given a chart on spacetime, we can view the Einstein tensor as 10
functions on
${\cal J}$.  In particular, the vacuum Einstein equations
$$
G_{ab}=0
$$
define 10 relations between the metric and its first two derivatives,
{\it
i.e.}, a subspace of ${\cal J}$.  Of course, any solution to the field
equations will
also solve all equations obtained by taking all derivatives of the
field
equations, and these equations provide additional relations in ${\cal
J}$. 
Thus
solutions to field equations define points on the {\it Einstein
equation
manifold}, ${\cal E}\in {\cal J}$, which is the set of points in
${\cal J}$
satisfying the
relations dictated by the Einstein equations and all of their
derivatives.  Of
course, it is not obvious that $\cal E$ does form a manifold, but
it can be
shown that ${\cal E}$ is in fact a smooth sub-bundle of ${\cal J}$
[6].  By
the way, do not confuse $\cal E$ with the space of solutions to
the Einstein equations.  Solutions to the Einstein
equations
define cross sections of ${\cal E}$, just as metric tensor fields
define
cross-sections of the (jet) bundle of metrics $E$ (${\cal J}$).  In
particular, unlike ${\cal E}$, the
space of solutions to the Einstein equations need not be a manifold
[10].

Conservation laws only need hold modulo the field equations, or as
one often says, ``on shell''.  What this means is that the total
exterior derivative of a $p$-form conservation law, which is a
$(p+1)$-form-valued function on ${\cal J}$, must vanish when
evaluated on
points in $\cal E$.  By the same token, any two $p$-form conservation
laws that are equal on shell should be considered physically
equivalent.  In effect, what this means is that $p$-form conservation
laws are determined by their restriction to $\cal E$. For these
reasons
it is often quite useful to have an explicit parametrization of $\cal
E$.  Such a parametrization can be found in [6,13] and has proven
useful in
a variety of applications, so we briefly summarize it here.

Because the Einstein equations only restrict the geometry of
spacetime, they do not place any restrictions upon the jet
coordinates $(x^i, g_{ij}, \Gamma^{(k)})$, $k=1,2\dots\ $.  The second
derivatives of the metric are restricted via the vanishing of the
Ricci tensor, or, equivalently, all traces of $Q^{(2)}$ must
vanish.  Remarkably, this pattern continues to all the higher
derivatives.  More precisely, the vacuum Einstein equations and all
of their derivatives completely fix the traces of the variables
$Q^{(k)}$ in terms of their completely trace-free parts, which we
shall denote by $\widetilde Q^{(k)}$:
$$
\widetilde Q_{ij},_{j_1\cdots 
j_k}=g_{ir}g_{js}\nabla_{(j_3}\cdots\nabla_{j_k}
R^{r\phantom{(j}s}_{\phantom{r}j_1\phantom{s} j_2)}
 - ({\rm all\ traces}), \quad
k=2,3,\ldots\ .\eqno(4.1)
$$
Thus the variables 
$$
(x^i, g_{ij}, \Gamma^{(1)},\Gamma^{(2)},\ldots,\widetilde
Q^{(2)},\widetilde Q^{(3)},\ldots)\in {\cal E}\eqno(4.2)
$$
parametrize points in $\cal E$, {\it i.e.}, are the freely specifiable
data at a point of an Einstein spacetime.  

The coordinates (4.2) on the 
equation manifold $\cal E$ can be interpreted in terms of a power
series
expansion 
of an Einstein metric.  More precisely, if we are trying to build an
Einstein 
metric by Taylor series we (i) specify the spacetime point $x^i$
around 
which the series is being developed, (ii) specify the metric 
components $g_{ij}$ at $x^i$, (iii) specify the variables
$\Gamma^{(k)}$; this 
fixes the coordinate system in which the metric components are being
built, (iv)
specify 
the variables $\widetilde Q^{(k)}$, which supplies the geometric
content of 
the Einstein metric.  {}from this freely specifiable data, the Einstein
equations can be used to construct all derivatives of the metric at
the chosen point, which defines the formal power series solution to
the field equations.  Of course, this procedure leaves open the
question of convergence of the series.

The parametrization (4.2) turns out to be somewhat unwieldy in
applications, 
primarily because of the need to remove so many traces.  A much more 
useful parametrization uses a spinor representation of the variables 
$\widetilde Q^{(k)}$ [6,7,11,13].  The spinor representation of
$\cal E$
that arises in 4 spacetime dimensions is as follows.  
Let $\Psi_{\scriptscriptstyle ABCD}$
and
$\overline\Psi_{\scriptscriptstyle 
A^\prime B^\prime C^\prime D^\prime}$ denote the Weyl spinors (see
[12] for definitions).  Fix a 
soldering form $\sigma_a^{\scriptscriptstyle AA^\prime}$ such that,
for a given $g_{ij}$,
$$
g_{ij}=\sigma_i^{\scriptscriptstyle
AA^\prime}\sigma_{j\scriptscriptstyle AA^\prime}.
$$
It can be shown that the variables $Q^{(k)}$ are uniquely
parametrized by 
the soldering form, the spinor variables
$$
\Psi^{(k)}\longleftrightarrow \Psi{}_{\scriptscriptstyle J_1\cdots
J_{k+2}}^{\scriptscriptstyle 
J_1^\prime \cdots J_{k-2}^\prime}
=\nabla_{\scriptscriptstyle (J_1}^{\scriptscriptstyle(
J_1^\prime}\cdots\nabla_{\scriptscriptstyle
J_{k-2}}^{\scriptscriptstyle 
J^\prime_{k-2})}\Psi_{\scriptscriptstyle J_{k-1}J_kJ_{k+1}J_{k+2})}
\eqno(4.3)
$$
and their complex conjugates $\overline\Psi{}^{(k)}$.
Thus we obtain a spinor parametrization of the Einstein equation
manifold  
in terms of
$$
(x^i,g_{ij},\Gamma^{(1)},\Gamma^{(2)},\dots, 
\Psi^{(2)},\overline\Psi{}^{(2)},\Psi^{(3)},\overline\Psi{}^{(3)},
\dots).\eqno(4.4)
$$
This spinor representation of $\cal E$ proved essential in the
classification of symmetries of the Einstein equations (see [11,13]
and
\S8 below).

\bigskip
\noindent{\bf 5. $p$-form conservation laws and local
cohomology.\hfill}

We are now ready to give a precise definition of a $p$-form
conservation law.  We consider a $p$-form $\omega[g]$, locally
constructed {}from the metric, to be a map {}from ${\cal J}$ into the
spacetime
$p$-forms depending upon a finite but arbitrary number of derivatives
of the metric.  We say that $\omega[g]$ is a {\it $p$-form
conservation law} if
$$
D\omega = 0 \qquad{\rm on}\ {\cal E}.
$$
Of course, as we have noted before, if $\omega[g]$ is an exact
$p$-form then it is of no interest.  We therefore say that a $p$-form
conservation law is {\it trivial} if there is a $(p-1)$-form
$\alpha[g]$, locally constructed {}from the metric and its derivatives
to some finite order, such that
$$
\omega[g]=D\alpha[g].
$$
Any two $p$-form conservation laws that differ by a trivial
conservation law on-shell will define the same conserved quantity;
consequently it is useful to identify any two conservation laws that
differ by an exact form on-shell:
$$
\omega\sim\omega^\prime
$$
if there exists $\alpha[g]$ such that
$$
\omega=\omega^\prime + D\alpha\quad{\rm on}\ {\cal E}.
$$

Thus we are lead to consider on-shell equivalence classes $[\omega]$
which
define the 
{\it local cohomology} of the field theory.  This cohomology can be
obtained by pulling back the Euler-Lagrange complex {}from ${\cal J}$
to $\cal E$ using the natural embedding of
$\cal E$ into ${\cal J}$.  Even if there is no cohomology in the
Euler-Lagrange complex associated to the jet bundle ${\cal J}$ ({\it
i.e.}, there are no topological conservation laws, see \S6), there
{\it can} be
cohomology ({\it i.e.}, $p$-form conservation laws) in the
Euler-Lagrange complex associated to the equation manifold $\cal E$.
This situation is somewhat analogous to the fact that, {\it e.g.}, all
closed forms are exact on ${\bf R}^3$, but this is not so on the
torus ${\bf T}^2\hookrightarrow{\bf R}^3$.  A better, rather
elementary example of this phenomenon, taken {}from field
theory, is as follows.  Let the field
theory of interest be a that of a scalar field $\varphi$ on an
$n$-dimensional spacetime.  It can be shown (see [14] and/or \S6
below)
that identically closed $p$-forms, with $0<p<n$ locally constructed 
{}from $\varphi$
are necessarily exact.  On the other hand, the following $(n-1)$-form
is closed on the equation manifold defined by the wave equation
$\nabla_a\nabla^a\varphi=0$:
$$
\omega[\varphi]=\epsilon_{a_1\dots a_n}\nabla^{a_n}\varphi\
dx^{a_1}\wedge
\dots \wedge dx^{a_{n-1}}.\eqno(5.1)
$$
But there is no $(n-2)$-form locally constructed {}from $\varphi$ whose
exterior derivative yields $\omega[\varphi]$ on shell. Analogous
remarks apply to the conserved form (2.2) for the charged scalar
field.

It is important to stress at this point that even if the de Rham
cohomology of the spacetime manifold is completely trivial, there can
still be interesting local, field theoretic cohomology. So, for
example, suppose the free scalar field theory, {}from which (5.1) is
derived, is formulated on ${\bf R}^n$.  If we evaluate the
$(n-1)$-form (5.1) (or (2.2) in the charged case) on any particular
solution $\varphi(x)$ of the field equations then the resulting
closed form 
$\omega(x)=\omega[\varphi(x)]$ on $M$ is guaranteed to be exact (as a
form on $M$) since all closed forms are exact on ${\bf R}^n$.  This
does not alter the fact that (5.1) defines a non-trivial cohomology
class, that is, a $p$-form conservation law, for the field theory. 
The triviality on $M$ of the form $\omega(x)$ does {\it not}
imply the triviality of the form $\omega[\varphi]$ on $\cal E$ since
the
$(n-2)$ form $\alpha(x)$ on $M$ which satisfies 
$\omega(x)=d\alpha(x)$
cannot be constructed locally {}from the field and its derivatives. 
For this reason, field theoretic cohomology is sensitive to
structures such as conservation laws without the need to specify
global data such as boundary conditions on solutions.
 
\bigskip
\noindent{\bf 6. Identically closed $p$-forms: cohomology of the free
Euler-Lagrange complex.\hfill}

Before investigating methods for computing closed $p$-forms locally
constructed {}from solutions to the field equations, it is worth
pausing to first understand {\it identically} closed $p$-forms, that
is, $p$-form conservation laws that arise irrespective of field
equations.  As we shall see, such $p$-form conservation laws reflect
topological properties of the bundle of fields.  Consequently, we
shall refer to equivalence classes of such conservation laws (modulo
exact forms) as {\it topological conservation laws}.

We now cite, without proof and somewhat informally, some basic
results concerning the origins of topological conservation laws. 
Again, for simplicity, we phrase the results in terms of our prime
example:  general relativity, but it is not hard to see how to
generalize these results to other field theories.

Our first result is that there is an isomorphism between the
cohomology classes of the Euler-Lagrange complex and that of the de
Rham complex on ${\cal J}$.  This result is proven, {\it e.g.}, in
[9] using standard techniques {}from homological algebra in the
variational bicomplex.  Next, it can be shown that the de Rham
cohomology of the jet bundle ${\cal J}$ is isomorphic to that of the
bundle
of metrics $E$.  Thus, modulo exact $p$-forms locally constructed
{}from a metric, all identically closed $p$-forms which are constructed
locally {}from a metric can be obtained by finding non-trivial
cohomology classes in $E$. The construction of a closed form on $M$
{}from a closed form on $E$ is simple to exhibit in local coordinates;
it amounts to pulling back the form on $E$ by a
cross-section.  Recall that coordinates on $E$ are $(x^i,g_{ij})$,
where $x^i$ are coordinates on spacetime.  Thus, for example, a
differential 1-form $\alpha$ on $E$ can be written as
$$
\alpha=A _i(x,g)dx^i + B^{ij}(x,g)(x,g) dg_{ij}.
$$
The corresponding form $\alpha[g]$ on $M$ is obtained by
$$
\alpha[g]=[A _i(x,g)+ B^{jk}(x,g)g_{jk,i}]dx^i.
$$
It is a simple exercise to show that if $\alpha$ is a closed form on
$E$, then $\alpha[g]$ is a closed form on $M$ locally constructed
{}from the metric.  It is also a simple exercise to generalize this
example to any $p$-form; one simply pulls back the form {}from $E$ to
$M$ using an arbitrary local section and reinterprets the result as
an element of the Euler-Lagrange complex.  The main result, then, is
that modulo exact forms all identically closed $p$-forms locally
constructed {}from a metric
arise this way.  Thus, topological conservation
laws for a metric theory can always be expressed via the metric and
its first derivatives only.  

If the bundle of fields is a vector bundle, then all
cohomology of the bundle comes {}from lifting closed forms {}from the
base space (the spacetime
manifold) [25]. These forms, of course, have nothing to do {\it per
se} with the field theory. Thus in this case there will be no
interesting topological
conservation laws ({\it cf.} [14]).  On the other hand, the bundle of
metrics is not a vector bundle, and there {\it do} exist topological
conservation laws for field theories based upon pseudo-Riemannian
metrics.

\bigskip
\noindent{\bf 7. Identically closed $p$-forms locally
constructed {}from a metric:  gravitational kink number.\hfill}

To compute topological conservation laws in metric theories of
gravity, we need to understand the cohomology of the bundle of
metrics $E$ over the spacetime manifold $M$.  I remind you that this
is not the infinite-dimensional space of metrics, but rather is the
finite dimensional bundle (${\rm dim}\ E=14$ when ${\rm dim}\ M= 4$)
whose cross sections are metric tensor fields.  In coordinates, a
point in $E$ is a pair $(x^i,g_{ij})$, where $x^i$ labels a point in
the spacetime manifold and $g_{ij}$ are components of a symmetric,
non-degenerate rank-2 tensor with the appropriate signature. We will
assume $M$ is orientable and the metric is Lorentzian, {\it i.e.}, it
has signature $(-+++\dots)$. 

It is shown by Steenrod [15] that the bundle of Lorentzian signature
metrics over an
$n$-dimensional spacetime manifold admits a deformation retraction to
a bundle $E^\prime$ over spacetime whose typical fiber is
diffeomorphic to the real projective space $RP^{n-1}$ (which can be
defined as the space of lines through the origin in ${\bf R}^n$), so
that
it is the cohomology of {\it this} bundle that we need to compute.  
The
origin of the real projective space is fairly easy to understand.  It
is a standard result {}from Lorentzian geometry (see e.g., [16]) that a
manifold admits a Lorentzian metric if and only if it admits a {\it
line element field} (or {\it direction field}), which is a continuous
assignment of a vector, up to a non-zero multiplicative factor, to
each point
of the manifold.  This line element field is, at each point of $M$, a
line in
the tangent space
to the point which lies inside the light cone.  To compute this line
in a coordinate chart at any given point, one simply finds the
eigenvector with negative eigenvalue of the matrix of components of
the metric in
that chart at the point of interest. With Lorentzian signature, the
resulting vector is unique up to a multiplicative factor.  Thus the
metric defines, at each point, an element of $RP^{n-1}$. In fact, it
can be shown that the metric defines a global mapping {}from spacetime
into $RP^{n-1}$.  The map is not canonical however; it depends upon
other data besides just the manifold and metric.  

It can be shown
[17] that the cohomology of $E^\prime$ comes {}from either the base
manifold or {}from the fibers.  The
cohomology of $M$ leads to conservation laws that have nothing to do
with the metric. Only the fiber cohomology leads to
interesting topological conservation laws in the sense that it leads
to conserved $p$-forms locally constructed {}from metric.  So we need
the cohomology
of $RP^{n-1}$, which is given by constants at degree zero and volume
forms at degree $n-1$ when $n$ is even.  When $n$ is odd, $RP^{n-1}$
is not orientable and the only cohomology classes arise at degree
zero, so that the only conserved quantities are the rather
uninteresting constant functions.  Thus, non-trivial topological
conservation laws arise when the
spacetime dimension is even, and are obtained by pulling back a volume
form on $RP^{n-1}$ to spacetime using the line element field
construction [17].  The resulting topological conservation law is of
degree $n-1$, and so corresponds to a topological current.  The
conserved quantity obtained by integrating the normal component of
this current over a
hypersurface is the ``kink number'' conservation law, first obtained
by Finkelstein and Misner in 1959 using homotopy considerations
[18].  In even-dimensional, orientable spacetimes the
kink number is the degree of the map {}from a hypersurface
to
$RP^{n-1}$ which is provided by the metric.  Somewhat more
figuratively, the
kink number measures the number of times a light cone tumbles as one
moves along the hypersurface.  This kinking of the light cone field
defines an integer.  A continuous deformation of the hypersurface
induces a continuous change in the metric evaluated on the surface,
which, however, cannot induce a continuous change in an integer. 
Thus the kink number is conserved.  It can be shown [18] that if the
kink number is non-zero then there will be no hypersurfaces that are
everywhere spacelike (in the homology class with non-zero kink
number).  

To summarize, in even
dimensional, orientable spacetimes, the kink number of Finkelstein
and Misner can be interpreted as arising {}from an
identically closed $(n-1)$-form locally constructed {}from the
metric.  Modulo exact forms, this is the only topological
conservation law for metric field theories.  Let me add that this kink
conservation law does not arise if the metric is Riemannian, since in
this case $E$ retracts to $M$ [15].  The
reason for this is clear:  the kink conservation law arises because
of the possibility of distinguishing a timelike direction at each
point of spacetime. This possibility does not arise with Riemannian
metrics.

\bigskip
\noindent{\bf 8. $(n-1)$-form conservation laws and
infinitesimal symmetries of the field equations.}

Now we consider the problem of finding and explaining $(n-1)$-form
conservation laws ($n$ is the spacetime dimension) associated to
field equations such as the Einstein equations.  Given a metric
on spacetime, such conservation laws are equivalent to the existence
of conserved currents via the Hodge duality operation.  Thus we are
on the familiar territory usually handled via Noether's theorem
[19].  Consequently, nothing I will reveal here is new,
field-theoretically speaking, except perhaps the presentation, which
is not so traditional in the physics literature.  My mode of
presentation is designed to jibe with an analogous treatment of
lower-degree conservation laws in subsequent sections, which 
{\it is} new.  Again, it is convenient to present the discussion in
the context of our prime example, the vacuum Einstein field equations.

So, consider a spacetime $(n-1)$-form locally constructed {}from the
metric which is closed when the vacuum Einstein equations hold.  This
is equivalent to the existence of a spacetime vector field $J^a[g]$ 
locally constructed {}from the metric which is divergence-free on-shell,
$$
\nabla_aJ^a[g] = 0\qquad{\rm on}\ {\cal E}.
$$
Here the covariant divergence is built {}from the total derivative in
the obvious way.  Because the Einstein equation manifold $\cal E$
is a submanifold of ${\cal J}$, it can be shown that this
conditional
equality is equivalent to the existence of tensor-valued functions,
$\rho_{ab}$, $\rho_{abc}$, $\ldots$, on ${\cal J}$ such that there is
an
{\it identity} on ${\cal J}$
$$
\nabla_aJ^a[g] = \rho_{ab}G^{ab} + \rho_{abc} \nabla^cG^{ab} + \dots\
,
\eqno(8.1)
$$
where the number of terms on the right-hand side of (8.1) depends
upon the derivative order of the metric appearing in
the current $J^a$.  Of course, the current is only defined by this
relation up to the possible addition of an identically conserved
current. For example,
$$
J_0^a[g] = \nabla_b S^{ab}[g],\qquad S^{ab}=-S^{ba},\eqno(8.2)
$$
is divergence-free identically, {\it i.e.}, it satisfies (8.1) with
all
the multipliers $\rho$ vanishing.  In the language of differential
forms, the conserved current $J_0^a$ is dual to an exact
$(n-1)$-form, and so is uninteresting.  Ignoring topological
conservation laws, which we have already enumerated for this theory,
the only ambiguity in $J^a$ is that associated with trivial
conservation laws (8.2).

The identity (8.1) can be considerably simplified since we are only
interested in the on-shell values of the current $J^a$.  For example,
note that the second term on the right hand side of (8.1) can be
written
as
$$
\rho_{abc}
\nabla^cG^{ab}=\nabla^c(\rho_{abc}G^{ab})-(\nabla^c\rho_{abc})G^{ab}.
\eqno(8.3)
$$
The first term in (8.3) can be absorbed into an on-shell trivial
redefinition of $J^a$.  The second term can be absorbed into a
redefinition of the multiplier 
$\rho_{ab}$.  It is easy to see that a similar trick works on all
higher-derivative terms in (8.1).  Thus, up to on-shell trivial
redefinitions of $J^a$, we can consider the simpler identity
$$
\nabla_aJ^a[g] = \rho_{ab}[g]G^{ab}.\eqno(8.4)
$$
In differential form language, this is the same as 
$$
D\omega[g] = \Lambda_{ab}[g]G^{ab},\eqno(8.5)
$$
where $\omega[g]$ is an $(n-1)$-form and $\Lambda_{ab}[g]$ is a
symmetric-tensor-valued $n$-form, both locally constructed {}from the
metric and its derivatives.  Our task is to find all identities of the
form (8.4) or (8.5).  Notice that the obvious integrability condition
for
(8.5),
$$
D(\Lambda_{ab}G^{ab})=0
$$
is identically satisfied since $\Lambda_{ab}G^{ab}$ is a spacetime
form of
maximal degree.    One might be tempted to think that (8.5)
must therefore always have solutions, which is of course false. 
Keep in mind that the equations (8.4-5) are differential equations on
${\cal J}$,
not on the spacetime manifold $M$.  The correct
integrability condition for (8.5) is that the Euler-Lagrange operator
annihilate 
the ``trivial Lagrangian $n$-form'' $\Lambda_{ab}G^{ab}$. We have
already indicated this fact when discussing the structure of the
Euler-Lagrange complex in \S3.
Now, one can compute in a straightforward, albeit lengthy, manner the
restrictions placed upon $\Lambda_{ab}$ or $\rho_{ab}$ by this
integrability condition.  Keep in mind, though, that {\it a priori}
the multipliers depend upon an arbitrary but finite number of
derivatives of the metric. A useful short cut, based upon the
relation of the Euler-Lagrange operator to the variational calculus,
is as follows.

Integrate both sides of (8.4) over a region $B\subset M$:
$$
\int_{B}\sqrt{|g|} \nabla_aJ^a[g] =\int_B\sqrt{|g|}\rho_{ab}G^{ab}.
\eqno(8.6)
$$
Because the integrand on the left-hand-side of (8.6) is a total
divergence, we get
$$
\int_{\partial B}\sqrt{|\gamma|} n_aJ^a[g]
=\int_B\sqrt{|g|}\rho_{ab}G^{ab},
\eqno(8.7)
$$
where $n_a$ is the unit normal to the boundary $\partial B$, which we
assume is non-null, and $\gamma_{ab}$ is the induced metric on
$\partial B$.
Consider a field variation $\delta g_{ab}$ of compact support.  For a
suitably ``large'' region $B$ the boundary $\partial B$ will be
outside the support of $\delta g_{ab}$.  The change in the left-hand
side of (8.7)
induced by this variation vanishes, so that we conclude
$$
\int_B\delta\left(\sqrt{|g|}\rho_{ab}G^{ab}\right)=
\int_B\left(\sqrt{|g|}G^{ab}\delta\rho_{ab}
+\rho_{ab}\delta(\sqrt{|g|}G^{ab})\right)=0
\eqno(8.8)
$$
for variations $\delta g_{ab}$ of compact support.  Because the
Einstein equations come {}from varying an action, {\it i.e.}, are the
Euler-Lagrange equations for some Lagrangian, the linearized
Einstein operator 
$$
L^{ab}(\delta g)=\delta(\sqrt{|g|}G^{ab})
$$
is formally self-adjoint (see, {\it e.g.}, [19,20]):
$$
\int_B\rho_{ab}L^{ab}(\delta g) =  \int_B\delta g_{ab}L^{ab}(\rho).
$$
Using (8.8) we find that
$$
\int_B\delta g_{ab}L^{ab}(\rho) = 0
$$
whenever $G_{ab}=0$ and $\delta g_{ab}$ is of compact support.  By
the fundamental theorem of the variational
calculus, we conclude that
$$
L^{ab}(\rho) = 0 \quad {\rm on}\ {\cal E}.
\eqno(8.9)
$$
To summarize:  every conserved current has an associated
solution $\rho_{ab}[g]$ of the linearized equations locally
constructed {}from the
fields and their derivatives.  Such a solution to the linearized
equations defines an infinitesimal {\it generalized symmetry}
transformation of the field equations in the sense that, if
$g_{ab}(x)$
is a solution to the Einstein equations, then to first order in
$\epsilon$ so is $g_{ab}(x) + 
\epsilon \rho_{ab}[g(x)]$.  We have arrived at one version, actually a
sort of converse, of Noether's theorem:  {\it Associated to every
conserved current of a Lagrangian field theory is a symmetry
transformation of the field
equations}.  
%
%extra
%
Topological conservation laws, such as the kink current, have
$\rho_{ab}=0$, which {\it is} a symmetry albeit a trivial one.
Note, however, that not every symmetry of the field
equations necessarily is associated to a conserved current via
(8.4).  A
symmetry of the field equations is necessary but not sufficient for
the existence of the conserved current.  As an elementary example,
consider a massless scalar field $\varphi$ satisfying the wave
equation
$$
\nabla^a\nabla_a\varphi=0.
$$
It is easy to see that for any constant $c$ the multiplier 
$$
\rho[\varphi]=c\varphi
$$
satisfies the linearized field equation (which is of course just the
wave equation again) when $\varphi$ satisfies its field equation.
This solution of the linearized equations reflects the scaling
symmetry of the wave equation.  However, there is no current locally
constructed {}from the field and its derivatives such that
$$
\nabla_a J^a=c\varphi\nabla_b\nabla^b\varphi,
$$
so that there is no conservation law associated with the scaling
symmetry.
Of course, a necessary and
sufficient condition is that the symmetry preserve a Lagrangian for
the field equations.  This is the basis of the Noether theory [19]. 
Typically, scaling symmetries of field equations do not preserve an
underlying Lagrangian.

Thus an integrability condition for the existence of a conserved
$(n-1)$-form as in (8.5) is the existence of a solution to the
linearized field equations locally constructed {}from the fields.  It
is not hard to see that this result is not specific to the Einstein
equations, but will arise for any variational system of field
equations.

%
%extra
%
\medskip
\noindent{\it Remark:\hfill}

For linear field equations,
schematically,
$$
L(\varphi)=0,
$$
every solution $\rho[\varphi]$ of the
field equations built locally {}from a solution $\varphi$
to the field equations has a corresponding conservation law.  This is
just an application of the identity
$$
\rho[\varphi]L(\varphi)-\varphi L(\rho) = \nabla_a J^a[\varphi],
$$
which holds for any linear differential operator $L$ for some vector
field $J^a$ locally constructed {}from $\varphi$ ( see {\it e.g.},
[20]).  This identity does not, however, guarantee that the
conservation law is non-trivial.  For example, the scaling symmetry
of the wave equation corresponds, in the above sense, to an {\it
identically} conserved
current, {\it i.e.}, an exact (hence trivial) $(n-1)$-form.

It is a routine (albeit sometimes challenging) affair to compute
generalized symmetries for field equations [19].  For example, the
generalized symmetries of the vacuum Einstein equations (in four
spacetime dimensions) have been classified as follows [11,13].  Let
$\rho_{ab}[g]$ be a tensor-valued function on ${\cal J}$ which
satisfies the
linearized Einstein equations on $\cal E$.  Then, modulo terms that
vanish on 
$\cal E$ ({\it i.e.}, on shell trivial solutions), there is a
constant $c$
and a vector field $X^a[g]$ locally constructed {}from the metric so
that
$$
\rho_{ab} = c g_{ab} + \nabla_{(a} X_{b)}.\eqno(8.10)
$$
The first term represents the scaling symmetry $g\rightarrow cg$
admitted by the vacuum equations, while the second term reflects the
diffeomorphism symmetry of the field equations.  Neither the scaling
symmetry nor the diffeomorphism symmetry are associated with
non-trivial conserved currents.  In detail, the scaling symmetry does
not preserve a Lagrangian, and the diffeomorphism symmetry corresponds
to an on-shell trivial conserved current:
$$
J^a = X_b G^{ab}.
$$
This strongly suggests, but does not quite prove, that there
are no non-trivial conservation laws for the Einstein equations
(aside {}from the kink conservation law).  The difficulty is that we
lack a strict 1--1
correspondence
between on-shell symmetries and conserved currents, the latter being
defined by off-shell identities (8.4).  It {\it is}
nevertheless true
that
there are no non-trivial, non-topological conservation currents for
the
vacuum Einstein equations, but the proof is rather involved: it
requires computing all conserved $p$-forms of
the linearized field equations, and the use of homological algebraic
constructions {}from the variational bicomplex [9] that are beyond the
scope of this survey.  A crucial ingredient in this classification of
conserved $(n-1)$-forms is a classification of conservation laws of
lower
form degree.  This is our next topic.

\bigskip\noindent
{\bf 9. Lower-degree conservation laws and
infinitesimal gauge symmetries of solutions to the field
equations.\hfill}

Our analysis of conserved $(n-1)$-forms, also known as conserved
currents, led us to an analysis of symmetries of the field equations,
by
virtue of integrability conditions of (8.5).  Our analysis of
lower-degree conservation laws will follow a similar strategy. 
Remaining in the context of our Einstein equations example, we begin
with
$$
D\omega[g]=\rho_{ab}[g] G^{ab} + \rho_{abc}[g]\nabla^cG^{ab} + \cdots 
+\rho_{abc\cdots r}[g]\nabla^c\cdots\nabla^r G^{ab},
\eqno(9.1)
$$
where $\omega[g]$ is a representative of a $p$-form conservation law
with $0\leq p<n-1$, and the multipliers $(\rho_{ab},
\rho_{abc},\dots)$ are tensor-valued $(p+1)$-forms locally
constructed {}from the
metric.  The integrability conditions for solutions to this equation
 are
$$
D\left(\rho_{ab}[g] G^{ab} + \rho_{abc}[g]\nabla^cG^{ab} + \dots
+ \rho_{abc\cdots r}[g]\nabla^c\cdots\nabla^r G^{ab}\right) = 0.
\eqno(9.2)
$$
The analysis of these conditions is rather different, and somewhat
more involved, than that which arose in the last section. Various
results on lower-degree conservation laws, obtained using a variety
of technologies, can be found in [1].  A general
theory of lower-degree conservation laws, and in particular equations
such as (9.2), will be given a detailed, rigorous presentation in a
forthcoming paper [2].  In these lectures I will simply illustrate a
typical set of results of the analysis via the Einstein equations in
four spacetime dimensions.  It is not too hard to infer the basic
features of the generalization of the gravitational results to
different
dimensions and other field theories.

Because the identities (9.2) must hold everywhere in $\cal J$, 
the integrability conditions reduce to a hierarchy of algebraic and
differential conditions on the multipliers $\rho$. These are the
lower-degree conservation law analogues of (8.9).   Using these
conditions it can be shown that, up to on-shell trivial redefinitions
of the conservation law $\omega$, that is, up to terms that are exact
when the field equations hold, the identity can be reduced to the form
$$
D\omega[g]=\rho_{ab}[g] G^{ab}.\eqno(9.3)
$$
Equation (9.3) is the lower-degree conservation law analog of
(8.5).  It should be kept in mind that the form $\omega$ and the
multiplier
$\rho_{ab}$ have been redefined in passing {}from (9.1) to (9.3).  In
order to keep the notation uncluttered, we retain the original
symbols.  It should be noted that, unlike the analysis of closed
$(n-1)$-forms, the ``integration by parts'' simplification of (9.1)
to (9.3) is {\it not} in general guaranteed for lower-degree
conservation laws.  The existence of such a simplification depends
upon details of the symbol of the field equations.  For gauge
theories of the type usually considered the simplification is
available.

We can now begin again with the analysis of the
integrability conditions for (9.3).  The
integrability conditions for these equations again involve algebraic
and differential conditions on the multipliers $\rho_{ab}[g]$.  The
algebraic conditions force
$\rho_{ab}$ to vanish if $\omega$ is a 0-form
or 1-form.  Thus we immediately conclude that there are only
topological conservation laws at form degree 0 and 1.   But there
are no interesting topological conservation laws at degree 0 and 1
(see \S7).  We thus see that the vacuum Einstein equations admit no
conserved 0-forms or 1-forms.

If $\omega$ is a 2-form, the same algebraic conditions that ruled out
conserved 0 and 1-forms imply the existence of a vector field
$X^a[g]$,
locally constructed {}from the metric and its derivatives, such that
$$
\rho_{ab}=X_{(a}\epsilon_{b)cde}dx^c\wedge dx^d\wedge dx^e.
$$
The remaining integrability conditions involve differential
conditions upon the 3-form multiplier $\rho_{ab}$ which are
equivalent to the total differential equation
$$
\nabla_{a}X_b + \nabla_b X_a = 0,  \quad {\rm on}\ {\cal E}.
\eqno(9.4)
$$
This formally looks like the Killing equation, 
and in a certain sense it is, but we
should be careful to understand it more precisely.  Eq.~(9.4) says
that
there is a vector field locally constructed {}from the metric and its
derivatives such that the symmetrized total covariant derivative
vanishes when the Einstein tensor and all its derivatives
vanish.  This is, of course, not quite the same as asking for a
Killing vector field, which is a spacetime vector field related to a
specific metric tensor field via the Killing equation.  Here we need
a map {}from the jet space of Einstein metrics into the space of vector
fields on $M$ that satisfies a particular differential equation in
${\cal J}$
when restricted to $\cal E$.  Roughly speaking, the existence of a
non-zero (on $\cal E$) solution to (9.4) would imply that every
solution
of the Einstein equations admits a Killing vector field.  This is of
course false.  ``Almost every'' solution of the vacuum Einstein
equations is devoid of symmetry [10]. Thus we expect that (9.4) has no
non-trivial ({\it i.e.}, non-vanishing on $\cal E$) solutions.  This
can be
proven directly using the adapted jet variables of \S4, but we will
not try to do it here (see [2] for details).  We conclude, again,
that $\rho_{ab}=0$ and
hence there are no conserved 2-forms associated to the vacuum field
equations.

Despite the negative result, we {\it have} learned something.  We see
that the Einstein equations {\it would} admit a 2-form conservation
law if every solution admitted a Killing vector field, more
precisely, if (9.4) admits non-zero solutions.  This result
generalizes to any Lagrangian field theory via the following ``rule
of thumb''. For a Lagrangian field theory in $n$-dimensions to admit
lower-degree conservation laws ({\it i.e.}, $p$-form conservation
laws with
$p<n-1$), the field theory must (i) be a gauge theory, {\it i.e.},
admit a
set of generalized symmetries which are constructed {}from arbitrary
functions, (ii) each solution of the field equations must admit a
{\it gauge symmetry}, {\it i.e.}, a gauge transformation that leaves
the solution
invariant. Thus, while a Lagrangian field theory must admit symmetries
of the Lagrangian in order to allow for conserved $(n-1)$-forms, it
must
admit gauge symmetries of solutions in order to allow for lower-degree
conservation laws.  This rule of thumb can be made quite rigorous and
precise.  For a derivation of this result in the context of the
BRST-antifield
formalism, and some illustrative examples, see [1].  A derivation of
these results in the spirit of the discussion above, a
variety of generalizations and examples, and a constructive procedure
for finding the conserved $p$-forms {}from solutions to the
integrability conditions will be provided in [2]. 

Let us revisit our Lagrangian field theory examples in light
of the above remarks. To begin, we expect that field theories which
do not admit gauge transformations will not admit lower-degree
conservation laws.  Thus, there are no lower-degree conservation laws
for the Klein-Gordon field.  This kind of no-go result can be made
quite general in terms of the rank of the symbol of the field
equations [2].  Roughly speaking, only gauge theories have a
sufficiently degenerate symbol to allow for lower-degree conservation
laws.
Next, the Maxwell equations for the 1-form $A$ admit
the gauge
transformation $A\rightarrow A + df$ for any function $f$ and thus
satisfy (i).  Moreover, $f={\rm constant}$ defines a gauge symmetry of
any solution of the Maxwell equations.  Thus (i) and (ii) are
satisfied in Maxwell theory and it can be shown that (2.4) is the
resulting lower-degree conservation law.  It can be shown that this
is the only lower-degree conservation law for the source-free Maxwell
equations.  Turning now to non-linear gauge theories, the vacuum
Einstein equations satisfy the criterion (i)
above---they admit
the diffeomorphism gauge transformation, but, as we have seen, fail
to satisfy (ii) and
thus admit
no lower-degree conservation laws.  Analogous results hold for the
Yang-Mills equations. The Einstein-Killing equations
(2.5)-(2.6) by construction satisfy both (i) and (ii) and this leads
to
the existence of lower-degree conservation laws such as (2.7) and
(2.8). 
Finally, the
two-dimensional dilatonic gravity models satisfy (i) because they
admit the diffeomorphism
gauge transformation
$$
g_{ab}\rightarrow \Psi^*g_{ab}\quad{\rm and}\quad \varphi\rightarrow
\Psi^*\varphi,
$$
where $\Psi\colon M\to M$ is a diffeomorphism and $\Psi^*$ denotes 
the corresponding pull-back.
Somewhat remarkably, every solution admits a
diffeomorphism gauge symmetry [3].  Explicitly, the following vector
field, which is locally constructed {}from the metric and dilaton field,
$$
X^a=\epsilon^{ab} \nabla_b \varphi
$$
satisfies the Killing equations when (2.10) is satisfied.  The
corresponding lower-degree conservation law is (2.11).

Finally, we introduce one more example of this phenomenon which is
relevant for our upcoming discussion of asymptotic conservation
laws.  Our example comes {}from the field theory obtained by
linearizing the Einstein equations about a given
solution. Let $(M,g)$ be an Einstein spacetime with cosmological
constant $\lambda$ (which can be set to zero if desired), 
$$
R_{ab}[g]=\lambda g_{ab}.\eqno(9.5)
$$
Let $h_{ab}$ be a symmetric tensor field.  The field equations for
$h_{ab}$ are the linearized Einstein equations
$$
-\nabla^c\nabla_c h_{ab} - \nabla_a\nabla_bh_c^c +
2\nabla^c\nabla_{(a}h_{b)c}-\lambda h_{ab}=0,\eqno(9.6)
$$
where all the geometric data arising in this equation (derivative
operator, etc.) are defined by the fixed Einstein metric
$g_{ab}(x)$.   
Let us consider the problem of determining what, if any, lower-degree
conservation laws are admitted by this linear field theory.  We begin
by
noting that this field theory admits the gauge transformation
$$
h_{ab}\longrightarrow h_{ab} + \nabla_{(a}X_{b)},
$$
where $X_a=X_a[g]$ is a covector-valued function on the jet space of
symmetric tensor fields and the derivative $\nabla$ is the total
covariant derivative.  Of course, this gauge transformation is just
the linearized remnant of the usual diffeomorphism transformation
(see (8.10))
that exists
for the full, non-linear Einstein equations.  Next, we note that if
$X^b(x)$ is a Killing vector field of $g_{ab}(x)$, then it defines a
gauge symmetry in the sense described above.  This suggests that the
linearized theory should admit lower-degree conservation laws if the
background metric admits Killing vector fields.  {}from an analysis of
the integrability conditions for lower-degree conservation laws in the
linearized theory, it can be
shown that the necessary and sufficient condition for a lower-degree
conservation law to exist is indeed that there exists a
Killing vector field of the fixed background metric $g_{ab}$. 
In particular, our analysis shows that the only non-trivial lower
degree
conservation law (in four spacetime dimensions) is of degree 2 and,
up to terms that are trivial on-shell, can be written as
$$
     \omega[h]
     = {1\over32 \pi}\epsilon_{ijhk}\big[ h^{li}\nabla_lX^j 
     - {1\over2}h^m_m \nabla^iX^j -X_l\nabla^ih^{lj} 
     +X^i(\nabla_l h^{jl} - \nabla^jh_m^m)]\, dx^h\wedge dx^k.  
\eqno(9.7)
$$
It is straightforward, albeit lengthy, to verify that $\omega[h]$ is
a closed 2-form provided $g_{ij}$ is an Einstein metric, $h_{ab}$
satisfies the linearized equations, and $X^a$ is a Killing vector
field.  Up to on-shell exact 2-forms, this is the only lower-degree
conservation law for the linearized theory.  The differential form
(9.7)
appears in [21] where it is used to study black hole entropy.  It can
also be used in asymptotic regions to define energy-momentum and
angular momentum of the gravitational field as we shall discuss in
the next section.

To summarize, based upon the rule of thumb presented above we can
see that, in general, a non-linear gauge theory will not admit
lower-degree
conservation laws because the generic solutions of the field
equations nominally admit no symmetries due to the intrinsically
non-linear nature of the field theory.  It is only for Abelian gauge
theories, such as Maxwell theory and linearized Einstein theory, that
the field equations are linear and do allow for gauge symmetries of
solutions and corresponding lower-degree conservation laws.

%
%extra
%
Having said this, one should wonder how the non-linear system of
field equations (2.9) and (2.10) describing the two-dimensional
dilaton gravity theory manage to allow for a lower-degree
conservation law.  The answer is that the lower-degree conservation
law (2.11) is really only associated to the field equation (2.10),
which, by itself, can be viewed as a linear equation for the dilaton
field $\varphi$ on a fixed spacetime $(M,g)$.

\bigskip
\noindent
{\bf 10.  Asymptotic conservation laws and local cohomology.}

In  
nonlinear gauge theories such as Yang-Mills  theory and general 
relativity, conserved quantities such as charge and energy-momentum
are computed {}{}from the limiting values of 
2-dimensional surface integrals in asymptotic regions.  A famous
example of this sort of conserved quantity is the ADM energy (2.12).
Such  {\it asymptotic  conservation laws\/} are most often 
derived by one of two, 
rather distinct,  methods. The oldest method relies upon the
construction
of identically conserved currents furnished by Noether's (second)
theorem and
subsequent extraction of
an appropriate ``super-potential'' to define a conserved surface
integral [21,22]. 
Unfortunately, to my knowledge, there is no intrinsic
field-theoretic criterion to select the appropriate
current:  for {\it any \/} field
theory there are infinitely many identically conserved currents that
can be
expressed as the divergence of a skew-symmetric super-potential,
modulo the field equations.  This
method of finding asymptotic conservation laws is thus somewhat {\it
ad
hoc}.  An alternative, more recent approach to finding asymptotic
conservation laws in gauge theories is based upon the Hamiltonian
formalism.  In this approach, asymptotic conservation laws arise as
surface term 
contributions to symmetry generators [23,21].  Given appropriate
asymptotic conditions, the generators of asymptotically non-trivial
gauge transformations are constructed as a sum of a volume integral
and
a surface integral in the asymptotic region.  When the field equations
are satisfied the volume integrals vanish.  Using the usual
Hamiltonian relation between symmetry generators and conserved
quantities, it follows that the surface integrals are asymptotic
conservation laws. The
Hamiltonian approach to finding asymptotic conservation laws
lacks the {\it ad hoc} flavor of
the super-potential formalism   but is somewhat
indirect:   to construct asymptotic conservation laws using this
method  one
must have the Hamiltonian formalism well in hand and 
 one must know {\it a priori} the general form of the
putative symmetry generator in order to find the appropriate surface
integral.

Here we discuss an interpretation of asymptotic conservation laws as
lower-degree conservation laws (typically $(n-2)$-forms in
$n$-dimensions) that are associated to the field equations and
asymptotic structure for the allowed solutions.  We have seen that,
generically, we expect no lower-degree conservation laws for for
non-linear gauge theories.  However, if the asymptotic structure of
the theory is such that, in an appropriate sense, the asymptotic
behavior of the fields is governed by the linearized theory, then the
asymptotic conservation laws can be viewed as lower-degree
conservation laws of the linearized theory.  Thus we are able to
establish that asymptotic conservation laws
for field theories can be viewed as arising {}{}from
(asymptotically) closed differential forms canonically associated to
the field
equations.   

To begin, let us give a general definition of an asymptotic
conservation law which appears sufficiently general to accommodate all
known examples. Denote by $M$ an asymptotic region of the
$n$-dimensional spacetime manifold and denote by
$\varphi^A$ the field of interest, which we assume is a cross section
of
some vector bundle or affine bundle (so that we can meaningfully take
the
difference between two fields).   We assume the asymptotic region $M$ 
is diffeomorphic to ${\bf R} \times ({\bf R}^{n-1}-C)$, where $C$ is
a 
compact set in ${\bf R}^{n-1}$.  The factor of ${\bf R}$ represents a
time
axis, and the 
${\bf R}^{n-1}-C$ factor represents space outside a compact
gravitating
system. 
Fix local  coordinates $(t, x^1, \dots, x^{n-1})$ 
in  the asymptotic region $M$.
We consider a fixed solution $\ovarphi^A$  to the field equations 
$\Delta_B[\varphi] =0$ 
and then, given a second solution $\varphi^A$, 
we set $h^A =\varphi^A- \ovarphi^A$. 
Let  $\omega[\ovarphi,h]$ be a spacetime $(n-2)$-form  
depending locally on the fields
$\ovarphi^A$ and $h^A$ and their derivatives.
We call  $\omega$ {\it an asymptotic conservation law
for the field equations $\Delta_B =0$ relative to the background
$\ovarphi$}  if, whenever $h^A$  satisfies the appropriate asymptotic
decay  conditions as $r = \sqrt{(x^1)^2 + (x^2)^2 +\dots +(x^{n-1})^2}
\to 
\infty$, the $p$-form $\omega[\ovarphi,h]$  satisfies 
$$
     \omega\sim {O}(1)\eqno(10.1)
$$
and
$$  
     D\omega\sim {O}(1/r).\eqno(10.2)
$$
Under these conditions, by Stokes theorem,
the limit 
$$
    Q[\ovarphi,h]=\lim_{r\to\infty} \int_{S_{(r,t_0)}}
\omega[\ovarphi,h],
\eqno(10.3)
$$
where $S_{(r,t_0)}$ is an $(n-2)$-dimensional sphere of coordinate
radius $r$ in
the 
$t=t_0$ hypersurface,  exists and is independent of $t_0$.  More
generally, granted the asymptotic conditions guarantee the vanishing
of the integral of the right hand side of (10.2) over the timelike
cylinder ${\bf R}\times S^2$ ``at infinity'', $Q$ in (10.3) is
conserved.

One way to construct asymptotic conservation laws is by, in effect,
mapping $(n-2)$-form conservation laws of the linearized theory into
the asymptotic structure of the theory.  A general construction is
described in [2,24].  Here, for reasons of brevity, I will just
sketch the idea in the context of our running example, the vacuum
Einstein equations in $4$ dimensions.   Let $\og_{ij}$ denote a fixed
fiducial Einstein
metric on the asymptotic region relative to which decay
conditions on solutions $g_{ij}$ of the Einstein equations
are specified.  In order to construct
asymptotic conservation laws for these equations, we first note that 
$\omega[h]$ in (9.7) is closed  by virtue of the identity
$$
	d\omega[h] = K_{ij} \nabla_k \alpha^{ijk} - \nabla_k K_{ij}\,
        \alpha^{ijk}  + L_{ij}[h]\,\beta^{ij},
\eqno(10.4)
$$
\noindent where $K_{ij} = \nabla_{(i} X_{j)}$,  $L_{ij}[h]$ is the 
linearization of $G_{ij} +\Lambda g_{ij}$ at $g=\og(x)$,
$$ 
\eqalign{
	\alpha^{ijk}
&        =  \bigl[h^{ij}\og^{kl}+ h^{kl}\og^{ij}
         -h^{jl}\og^{ik} - h^{ik}\og^{jl} +
          (\og^{ik}\og^{jl} -\og^{ij}\og^{kl}) h^m_m] \, \Omega_l
\cr
         \beta^{ij}
&        = 2 X^{(i}\og^{j)l}\, \Omega_l,\qquad{\rm and}\qquad
         \Omega_l
       = {1\over{3}}\varepsilon_{abcl}\,dx^{a}\wedge dx^b\wedge dx^c.}
$$
Note that we are now denoting the fixed, fiducial metric in the
asymptotic region as $\og$, and
$$
\nabla_a\og_{bc}=0.
$$
The key point now is to re-interpret the identity (10.4) in terms of 
asymptotic structure.  Define $h_{ij}=g_{ij}-\og_{ij}$.
By (10.4) and Stokes theorem we  now have 
the following existence principle for asymptotic conservation laws:
Granted that 
the asymptotic behavior of $X^i$, $h_{ij}$,  $K_{ij}$ and $L_{ij}[h]$
are
such that (10.5), shown below, is finite and the integrals of the
right-hand side of
(10.4) vanish asymptotically, then
$$
Q[\og,h]=\lim_{r\to\infty} \int_{S_{(r,t)}}
\omega[\og,h],\eqno(10.5)
$$ 
is conserved, that is, $Q[\og,h]$ 
is independent of $t$.
In particular, $Q$ is conserved
whenever $X^i$ is a Killing vector of the background $\og_{ij}$
and the decay conditions are such that the linearized
Einstein equations  for $h_{ij}$  are satisfied at an
appropriate rate at infinity.  

For example, when $n=4$,  
$\og_{ij}=\eta_{ij}$ and $X = \partial/\partial t$, 
a  straightforward calculation shows that
 (10.5) reduces 
to the ADM energy
$$
     Q[\eta,h]=
\lim_{r\to\infty}{1\over16\pi}\int_{S_{(r,t)}}\varepsilon_{bcd}
     (\partial_a h^{ab} - \partial^b h^{aa}) dx^c \wedge dx^d, \quad
     a,b,c,d =1,2,3,
$$
which is conserved given asymptotically flat fall-off conditions,
{\it e.g.},
$$
\eqalign{
h_{ij}&=O(1/r)\cr
h_{ij,k}&=O(1/r^2)\cr
h_{ij,kl}&=O(1/r^3).}
$$

This method of constructing asymptotic conservation laws has
a number of desirable field-theoretic properties.  

\medskip

\itemitem{\it (1)} First and foremost, it yields results in
agreement with those obtained using Hamiltonian methods
in the asymptotically flat and
asymptotically anti-De Sitter contexts.  In particular, the full
complement of conservation laws available for these asymptotic
structures can be obtained by letting $X$ range over the Killing
vectors of $\og$ in each case.

\medskip

\itemitem{\it (2)} The construction is manifestly
coordinate independent in the sense that the conservation laws are
obtained by integrating spacetime differential forms, such as (9.7),
which are defined in a coordinate independent manner.

\medskip

\itemitem{\it (3)} The charge $Q$ is unchanged by the substitution
$h_{ab}\to h_{ab}+\nabla_{(a}V_{b)}$ for any spacetime
co-vector field $V_a$.  This is crucial for establishing the ``gauge
invariance'' of the conservation law.

\medskip

\itemitem{\it (4)}  The construction readily generalizes to
other field theories, for example,
it is straightforward to extend our results to  the
Einstein-Yang-Mills equations, string-generated gravity models
(Lovelock gravity), etc.

\medskip

\itemitem{\it (5)} The construction is not tailored to any
 specific asymptotic structures.  In principle one  can  use the
fundamental identity (10.4) to find asymptotic conservation laws
for a wide variety of background metrics $\og$ and
fall-off conditions and, conversely, to analyze which
fall-off conditions admit conservation laws.  This feature also
arises with conserved currents:  Noether's theorem gives a formula
for a divergence-free vector field, independently of any boundary
conditions.  Whether or not the conserved current actually defines a
conserved quantity 
depends upon
boundary conditions.

\medskip

\itemitem{\it (6)}  The $(n-2)$-form $\omega[h]$
is constructed
directly {}from the field equations  with no reference made
to the Bianchi identities of the theory or to any Lagrangian
or Hamiltonian.  A general formula for $\omega[h]$ appropriate to
second-order field equations can be found in [24]. 
\bigskip

Thus asymptotic conservation laws can be viewed as a manifestation of
local cohomology associated to the field equations and asymptotic
conditions that are imposed.

\bigskip
\noindent{\bf 11. Slogans.}

I would like to conclude this introductory survey with a few slogans
that, to some extent, summarize the main ideas we have discussed.  As
with all slogans, they should be taken with a grain of salt.  Slogans
are no substitute for rigorous theorems, but they do help with the
intuition.  

The theme we have encountered is an old one: symmetries and
conservation laws.  We have seen that conserved currents depend upon
symmetries of the field equations for their existence.  In
particular, the lack of adequate symmetries of the vacuum Einstein
equations prevents the existence of conservation laws arising as
volume integrals of local densities.  Conserved quantities can also
be obtained by integrating locally constructed quantities over
lower-dimensional submanifolds.  The existence of such conservation
laws depends upon the existence of gauge transformations that
preserve solutions, a property we called a ``gauge symmetry''.  Thus,
while conserved $(n-1)$-forms are tied to symmetries of field
equations,
conserved forms of lower degree are tied to gauge symmetries of
solutions
to the field equations.  In particular, because almost all solutions
of the vacuum Einstein equations do not admit a symmetry, it follows
that there are no lower-degree conservation laws for the vacuum
equations.  These results for the vacuum field equations of general
relativity are quite typical of non-linear gauge theories.  Finally,
given appropriate asymptotic structure, it is possible for the
solutions to the field equations to admit an asymptotic symmetry. 
This leads to the existence of asymptotic conservation laws via
asymptotically closed $(n-2)$-forms.  

Thus a cornerstone of physics, the theory of symmetry and
conservation laws, is
neatly described via local cohomology in field theory.

\vfill\eject
\noindent
{\bf Acknowledgments}

I would like to express my gratitude to the organizers of the Second
Mexican School on Gravitation and Mathematical Physics for inviting
me to this stimulating conference in such a wonderful setting.  I
would also like to thank R. Capovilla and D. Sudarsky for a
colaboraci\' on musical.  This
work was supported in part by a grant {}from the National Science
Foundation
(PHY96-00616).

\bigskip\bigskip
\noindent{\bf REFERENCES}
\bigskip
\oneandathirdspace

\noindent
[1]. A.~M.~Vinogradov, ``The C-spectral sequence, Lagrangian
formalism and conservation laws I,II'', {\it J. Math. Anal. Appl.}
{\bf 100}, 1
(1984);
T.~ Tjujishita, ``Homological method of computing invariants of
systems of differential equations'', {\it Diff. Geom. Appl.} {\bf 1},
3 (1991);
;
R.~Bryant and P.~Griffiths, ``Characteristic cohomology of
differential systems I'', {\it J. Am. Math. Soc.} {\bf 8}, 507
(1995);
G. Barnich, F. Brandt, and M. Henneaux, {``Local BRST cohomology in
the antifield formalism, I,II'', \it Commun. Math. Phys.}
{\bf 174}, 57, 93 (1995).
\medskip\noindent
[2].  I.~M.~Anderson and C.~G.~Torre, ``$p$-Form Conservation Laws in
Field Theory'', Utah State University Preprint (1997).
\medskip\noindent
[3].  See, for example, J. Gegenberg, G. Kunstatter and D.
Louis-Martinez, ``Observables for two-dimensional black holes'', {\it
Phys. Rev.} D{\bf 51}, 1781 (1995).

\medskip\noindent
[4].  R. Arnowitt, S. Deser, and C. Misner, ``The Dynamics of General
Relativity'', in {Gravitation: An Introduction to Current Research},
ed. L. Witten (Wiley, New York, 1962).

\medskip\noindent
[5].  D. Saunders, {\it The Geometry of Jet Bundles}, (Cambridge
University Press, Cambridge, 1989).

\medskip\noindent
[6]. I.~M.~Anderson and C.~G.~Torre, ``Two Components Spinors and
Natural Coordinates for the Prolonged Einstein Equation Manifolds'',
{\sl Utah State University Technical Report}, 1994.

\medskip\noindent
[7]. R. Penrose, ``A spinor approach to general relativity'', 
{\it Ann. Phys.} {\bf 10}, 171 (1960).

\medskip\noindent
[8].  T.~Y.~Thomas, {\it Differential Invariants of Generalized
Spaces}, (Cambridge University Press, Cambridge, 1934).

\medskip\noindent
[9].  I.~M.~Anderson, ``Introduction to the variational bicomplex'', 
{\it Contemp. Math.} {\bf 32}, 51 (1992).

\medskip\noindent
[10].  A. Fischer and J. Marsden, ``The initial value problem and the
dynamical formulation of general relativity'', in {\it General
Relativity: An
Einstein Centenary Survey}, ed. by S. Hawking and W. Israel,
(Cambridge University Press, Cambridge, 1979).

\medskip\noindent
[11].  I.~M.~Anderson and C.~G.~Torre, ``Symmetries of the Einstein
equations'', {\it Phys. Rev. Lett.} {\bf
70}, 3525 (1993).

\medskip\noindent
[12].  R. Penrose and W. Rindler, {\it Spinors and Spacetime, Vol. 1},
(Cambridge University Press, Cambridge, 1984).

\medskip\noindent
[13].  I.~M.~Anderson and C.~G.~Torre, ``Classification of local
generalized symmetries of the vacuum Einstein equations'', {\it
Commun. Math. Phys.} 
{\bf
176}, 479 (1996).

\medskip\noindent
[14].  R.~M.~Wald, ``On identically closed forms locally constructed
{}from a field'', {\it J. Math. Phys.} {\bf 31}, 2378 (1990).

\medskip\noindent
[15].  N. Steenrod, {\it The Topology of Fiber Bundles}, (Princeton
University Press, Princeton, 1951).

\medskip\noindent
[16].  S. Hawking and G. Ellis, {\it The Large Scale Structure of
Spacetime}, (Cambridge University Press, Cambridge, 1973).

\medskip\noindent
[17].  C.~G.~Torre, ``Some remarks on gravitational analogues of
magnetic charge'', {\it Class. Quantum Grav.} {\bf 12}, L43 (1995).

\medskip\noindent
[18].  D. Finkelstein and C. Misner, ``Some new conservation laws'',
{\it Ann. Phys.} {\bf 6}, 320
(1959).

\medskip\noindent
[19].  P. Olver, {\it Applications of Lie Groups to Differential
Equations}, (Springer, New York, 1989).

\medskip\noindent
[20].  B. DeWitt, {\it Dynamical Theory of Groups and Fields},
(Gordon and Breach, New York, 1965).

\medskip\noindent
[21].  V. Iyer and R. M. Wald, ``Some properties of Noether charge
and a proposal for dynamical black hole entropy'', 
{\it Phys. Rev.} D {\bf 50}, 846
(1994).

\medskip\noindent
[22].  See, {\it e.g.}, J.~N.~Goldberg, ``Invariant transformations,
conservation laws, and energy-momentum'', in {\it General Relativity
and Gravitation: One Hundred Years after the Birth of Albert
Einstein}, ed. A. Held, (Plenum, New York, 1980); J.~ Katz, J.~Bicak,
and D.~Lynden-Bell, ``Relativistic conservation laws and integral
constraints for large cosmological perturbations'', to appear in the
May 15, 1997 issue of {\it Physical Review} D.

\medskip\noindent
[23].  T.~Regge and C.~Teitelboim, ``Role of surface integrals in the
Hamiltonian formulation of general relativity'' {\it Ann. Phys.} {\bf
88}, 286
(1974).

\medskip\noindent
[24].  I.~M.~Anderson and C.~G.~Torre, ``Asymptotic conservation laws
in classical field theory'', {\it Phys. Rev. Lett.} {\bf
77}, 4109 (1996).

\medskip\noindent
[26].  R.~Bott and L.~Tu, {\it Differential Forms in Algebraic
Topology}, (Springer, New York, 1982).

\bye